\documentclass[aps,pre,superscriptaddress,floatfix,twocolumn,showpacs]{revtex4-1}

\usepackage{amsmath}
\usepackage{amsfonts}
\usepackage{amssymb}
\usepackage{graphicx}
\usepackage{bm}
\usepackage{color}
\usepackage{hyperref}
\usepackage{cases}
\usepackage{ulem}
\usepackage{multirow}
\usepackage{braket}

\usepackage{bm}

\def\stacksymbols #1#2#3#4{\def\theguybelow{#2}
	\def\verticalposition{\lower#3pt}
	\def\spacingwithinsymbol{\baselineskip0pt\lineskip#4pt}
	\mathrel{\mathpalette\intermediary#1}}
\def\intermediary#1#2{\verticalposition\vbox{\spacingwithinsymbol
		\everycr={}\tabskip0pt
		\halign{$\mathsurround0pt#1\hfil##\hfil$\crcr#2\crcr
			\theguybelow\crcr}}}

\begin{document}

\preprint{APS/123-QED}

\title{Narrow bandwidth Q-switched Erbium-doped fiber laser based on dynamic saturable absorption filtering effect}

\author{Zengrun Wen$^{1,2}$}
 \author{Kaile Wang$^{1,2}$}
 \author{Shuangcheng Chen$^{1,2}$}
 \author{Haowei Chen$^{1,2}$}
 \author{Xinyuan Qi$^{3}$}
 \author{Baole Lu$^{1,2}$}
 \email{lubaole1123@163.com}
 \author{Jintao Bai$^{1,2}$}%
 \email{baijt@nwu.edu.cn}
\affiliation{%
 $^1$State Key Laboratory of Photoelectric Technology and Functional Materials, International Collaborative Center on Photoelectric Technology and Nano Functional Materials, Institute of Photonics and Photon-technology, Northwest University, Shaanxi, Xi'an 710069, China\\
 $^2$Shaanxi Engineering Technology Research Center for Solid State Lasers and Application, Provincial Key Laboratory of Photo-electronic Technology, Northwest University, Shaanxi, Xi'an 710079, China\\
 $^3$School of Physics, Northwest University, Xi'an 710069, China
}%


\date{\today}

\begin{abstract}
We proposed a narrow spectral bandwidth Erbium-doped fiber (EDF) laser Q-switched by a homemade saturable dynamic induced grating (SDIG) which is introduced via reforming the structure of a fiber saturable absorbers FSA with a piece of EDF and a fiber Bragg grating. The SDIG integrates both saturable absorption and spectral filtering effect simultaneously, which was confirmed through theoretical analysis and experimental results for the first time, to the best of our knowledge. Further study verified that the spectral width of the Q-switched emissions is decided by the length of the SDIG and the input power of the pump source. The Q-switched pulse with the narrowest spectral width of about 29.1 pm achieved in this work is the narrowest bandwidth pulse in the domain of the FSA Q-switched fiber lasers when the length of SDIG and pump power are 20 cm and 250 mW, respectively. Our method provides a simple way to obtain the Q-switched pulses with narrow bandwidths, which have promising applications for nonlinear frequency conversion, Doppler LIDAR and coherent beam combinations.

\end{abstract}

\maketitle

\section{Introduction}

To realize pulsed emission in fiber lasers, Q-switching is one of the preferred technology to generate short and high energy pulses which are widely employed in optical communications, industrial processing, sensing, medicine and spectroscopy, etc. \cite{chenieee20}. Besides, nonlinear frequency conversion \cite{peremansol}, Doppler LIDAR \cite{ouslimani} and coherent beam combinations \cite{heoe14,zhouieee15} require that narrow bandwidths of the short pulses to elevate the conversion efficiency, measurement accuracy and beam quality. Generally, a Q-switching and a band-limited element are both necessary to achieve, separately, a Q-switched pulse emission and a narrow spectral bandwidth effect. On the one hand, active modulators with external signals (such as an acoustic optical modulator or a piezoelectric actuator \cite{lees32,Posada2017,Kaneda2004}) and passive saturable absorbers (e.g., semiconductor saturable absorption mirrors and two-dimensional materials) have both been exploited to obtain Q-switched operation \cite{Tse2008,Li2017Mode,lipr6,Yao2017Graphene}; On the other hand, bandpass filter, phase-shifted FBG and multimode interference filter \cite{Tse2008,Chakravarty2017,Popa2011} are also employed to narrow the bandwidth. Besides,  some configurations based on the spectral narrowing effect (e.g., suppressing ASE gain self-saturation and coupled interference filtering) have also been adopted to achieve narrow spectra \cite{Yao19oe,Anting2003}.  However, one has to face such a fact that such separated functions usually result in highly complex laser cavity with a pretty low reliability. In the last decade, a highly integrated reliable saturable absorber filter with both of saturable absorption and spectral filtering in one device was achieved by forming a filter in an unpumped (without 975 nm pump light) rare-earth-doped fiber \cite{poozeshjlt,yehoe15,yehlpl4}. However, these saturable absorber filters were commonly used to realize continuous-wave narrow bandwidth lasing because it is difficult for the rare-earth-doped fibers to meet the FSA Q-switching criterion due to their small absorption cross-sections and low doping concentrations in the corresponding radiation bands \cite{tsaioe}. Tasi, et. al. proposed a method of {\it{mismatch of mode field areas}} to make the unpumped EDF satisfy the Q-switching criterion $C_q>1$ or even $C_q>1.5$ \cite{oltsai}, but the spectral filtering and narrow bandwidth output were not involved in the laser. 

In this work, we proposed a method to achieve an SDIG by inserting a segment of unpumped EDF  between a circulator and a fiber Bragg grating (FBG). Theoretical analysis and experimental observations confirmed both the saturable absorption and spectral filtering can be realized simultaneously with such an SDIG. Further investigation showed the FSA Q-switching criterion in our laser can be degenerated to $C_q=1$ due to the spectral filtering of the SDIG. In addition, the spectral width of the Q-switched pulses can be easily modulated by the length of the SDIG and the input pump power. The proposed configuration is quite efficient to generate the Q-switched pulses with narrow bandwidths.

\section{Experimental configuration}
\begin{figure}[h!]
	\centering\includegraphics[width=8.5cm]{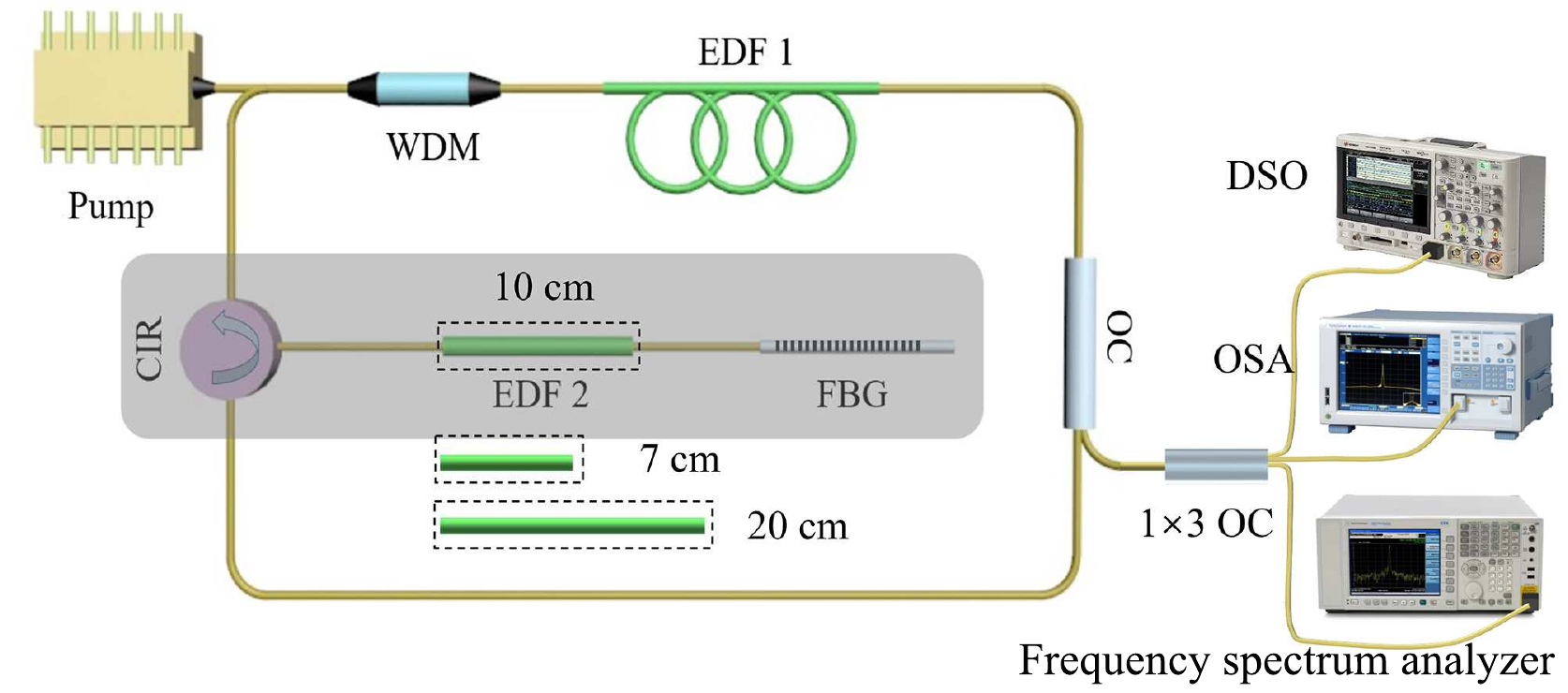}
	\caption{The schematic diagram of the all-fiber Q-switched laser. Inside the gray box is the SDIG.}
	\label{fig1}
\end{figure}
\noindent The architect of the proposed all-fiber Q-switched laser is depicted in Fig. \ref{fig1}. In the cavity, two pieces of EDFs (Liekki, Er110-4/125) are utilized for gain medium (with a length of 50 cm) and SDIG, respectively. All the components are directly connected by single-mode fibers (SMF-28e) and the core/inner cladding diameters of the EDFs and SMFs are 4/125 $\mu$m and 9/125 $\mu$m, respectively. The gain medium is pumped by a pigtailed diode laser emitting continuous wave at 975 nm through a 980/1550 nm wavelength division multiplexing (WDM). When the light goes through a 30/70 optical coupler (OC), 30$\%$ of the energy outputs and 70$\%$ continues to propagate in the cavity. Then, a circulator (CIR) with three ports and a reflective FBG (98$\%$ reflectivity at the central wavelength of 1550 nm) with the 3 dB bandwidth of 0.5 nm controls the light from port 1 to port 2, through the EDF2, reflected by the FBG, back to port 2 and port 3. Finally, the light enters into the WDM and finishes one roundtrip. The $\sim$10.3-m-long all-fiber cavity is compact and misalignment free, and all the components are commoditized. For measuring the output pulse, a real-time digital storage oscilloscope (DSO, Agilent Technologies, DSO9104A) with a bandwidth of 2.5 GHz, an optical spectrum analyzer (OSA, YOKOGAWA, AQ6370C) and a frequency spectrum analyzer (Agilent Technologies, N9000A) are employed to monitor the pulse trains, optical spectra and radio frequency signals, respectively.

\begin{figure}[h!]
	\centering\includegraphics[width=8.5cm]{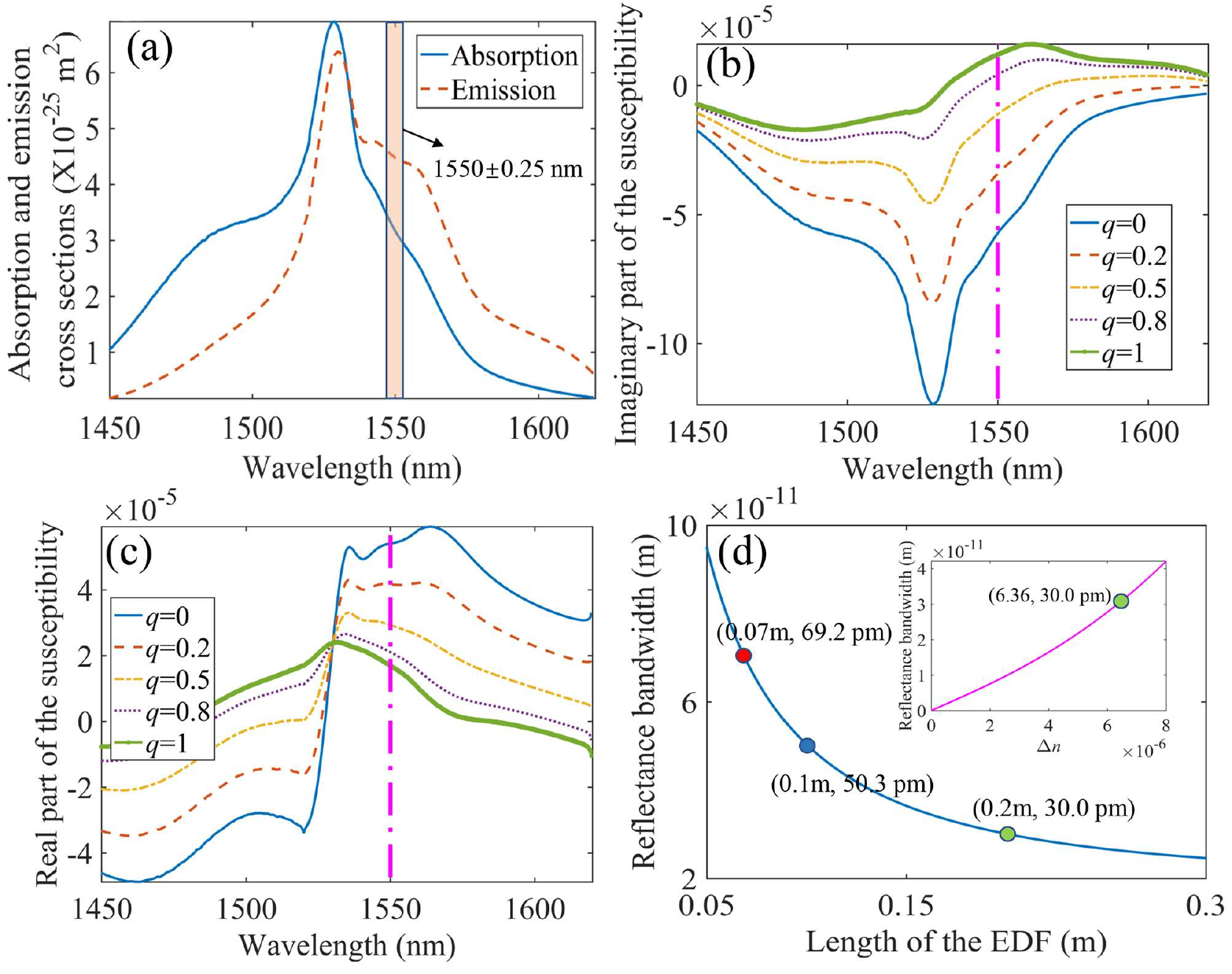}
	\caption{Characteristics of the SDIG. (a) Absorption and emission cross-sections of EDF; (b) and (c) imaginary and real part of the susceptibility versus normalized pump power from $q=0$ to $1$, respectively; (d) reflect bandwidth of the SDIG with respect to the length of EDF and the refractive rate change $\Delta n$ (inset).}
	\label{fig2}
\end{figure}
\noindent It was demonstrated that a fiber with high concentration of active ions can be used as a saturable absorber for generating a pulsed regime due to the ion clusters induced nonradiative transition~\cite{kurkovqe2010}. The doping concentration of the EDF in this work is $\rho=8\times 10^{25}\quad\rm ions/m^3$, which will be verified to be enough for a FSA in the experiment. In the cavity, the coherence of two counter-propagating lasing forms a standing-wave field between the circulator and the FBG. When the EDF2 absorbs the standing-wave field energy, a period spatial refractive index distribution is induced in the EDF2 due to the spatial selective saturation of the transition between the ground state and the excited state \cite{stepanovjpd}. Thus the SDIG is achieved. Figure. \ref{fig2}(a) depicts the absorption and emission cross-sections of the EDF2. The pink region represents the optical spectrum at 1550$\pm$0.25 nm which is limited by the FBG. In this region, the emission cross-section is larger than the absorption cross-section, and, nevertheless, there is saturable absorption characteristic in the EDF. The result shows inconformity with the opinion that the absorption cross-section should be larger than the emission cross-section \cite{tsaioe,oltsai}. Besides, in this setup, the saturable absorber Q-switching criterion is degenerated to $C_q=1$. We contribute that the grating region of the SDIG reflects the light step by step, and little energy reaches the back part of the SDIG. Thus, the back part of the SDIG still offers saturable absorption effect for a given power which is saturate for the FSAs without spectral filtering. In other words, the spectral filtering expands the Q-switching condition of the EDF. In the EDF2, the erbium ion transition occurs between energy levels $^4\rm I_{15/2}$ and $^4\rm I_{13/2}$ if the incident light is limited in 1550$\pm$0.25 nm region. Under this circumstance, the EDF can be regarded as a two-level system. Once the EDF2 absorbs light, the electric field of the light will result in the change of the susceptibility whose imaginary part $\chi''(\omega)$ is related to the absorption and emission cross-sections $\sigma_a$, $\sigma_e$ and atomic populations densities $N_1$, $N_2$. $\chi''(\omega)$ can expressed as \cite{desurvire}
\begin{equation}
-\chi''(\omega)=\frac{n_{eff}c}{\omega}[\sigma_e(\omega)N_2-\sigma_a(\omega)N_1],
\end{equation}
where $n_{eff}=1.46$ is the refractive index of the EDF and $c$ represents the light speed in vacuum. The relationship of the real and imaginary parts of the atomic susceptibility is expressed by  Kramers-Kronig relation (KKR)
\begin{equation}
\chi'(\omega)=\frac{1}{\pi}P.V.\int_{-\infty}^{+\infty}\frac{\chi''(\omega')}{\omega'-\omega}\rm d\omega',
\end{equation}
where $N_1=\rho_0/(1+q)$ and $N_2=\rho_0q/(1+q)$ describe the population densities at the two energy levels and $q$ is the normalized input power. $q=0$ and $q=1$ represent the EDF with no input power and saturation state, respectively. Figures \ref{fig2}(b) and (c) depict the imaginary and real parts of the susceptibility with different $q$. When $q$ is increased, $\chi''(\omega)$ enlarges and the absorption rate of the EDF reduces gradually. Meanwhile, the reduced $\chi'(\omega)$ reflects the decrease in refractive index change through $\delta n(\omega)=(\Gamma_s/2n)\chi'(\omega)$. In the EDF, the overlap factor $\Gamma_s=0.5$. From Fig. \ref{fig2}(c), the refractive index change at 1550 nm is calculated as $2.89\times10^{-6}<\delta n<9.25\times10^{-6}$, corresponding to a maximum refractive index difference $\Delta n$ of the EDF of $6.36\times10^{-6}$.

Inside the unpumped EDF, the formed DIG is considered as a Bragg reflective grating \cite{stepanovjpd}. Thus the FWHM bandwidth is described by \cite{zhangoe16}
\begin{equation}
\Delta \lambda=\lambda\kappa\sqrt{(\frac{\Delta n}{2 n_{eff}})^2+(\frac{\lambda}{2n_{eff}L_g})^2},
\end{equation}
where $\lambda$ and $L_g$ are the central wavelength of light and the length of EDF, respectively. $\kappa=2\Delta n/(\lambda n_{eff})$ is the coupling coefficient of the DIG. The reflective bandwidth versus $L_g$ and $\Delta n$ is shown in Fig. \ref{fig2}(d). The reflectance bandwidth decreases as the EDF lengthens and $\Delta n$ (related to the input power of pump source) becomes small. The marks represent the lengths and $\Delta n$ of the used EDFs in this work. The reflective bandwidths $\Delta \lambda$ are calculated as 69.2 pm, 50.3 pm and 30.0 pm with the EDF lengths of 7 cm, 10 cm and 20 cm, respectively. Apparently, the saturable absorption and spectral filtering can both be achieved for the SDIG, thus it can be used as a narrow bandwidth SA.

\section{Experimental results}
\begin{figure}
	\centering\includegraphics[width=8.5cm]{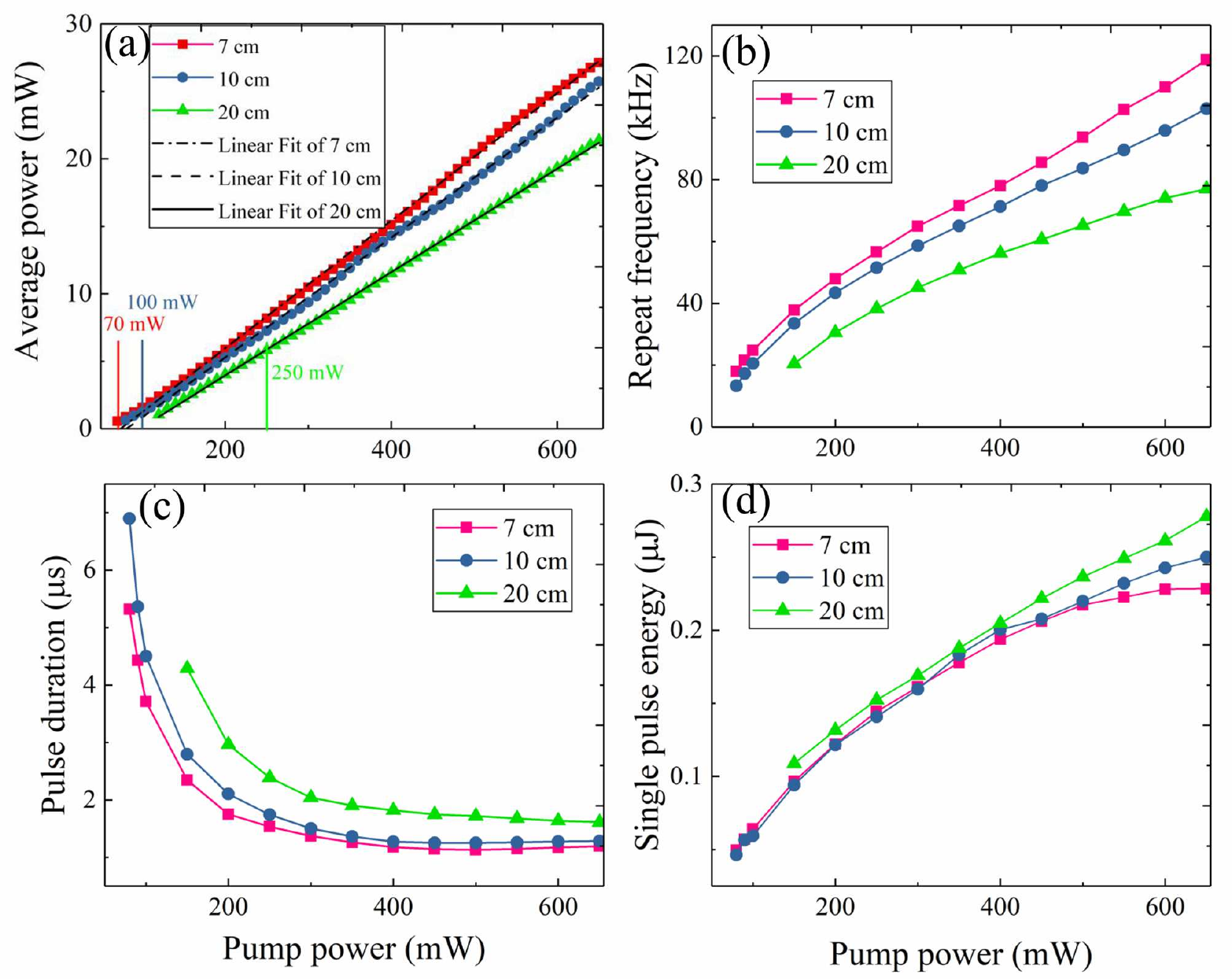}
	\caption{Pulse characteristics including (a) average powers, (b) repeat frequencies, (c) pulse durations and (d) single pulse energies versus pump power.}
	\label{fig3}
\end{figure} 

\noindent In our experiment, we modulated the pump power from 1-650 mW and measured the Q-switching performances in terms of the average powers, repeat frequencies, pulse durations and single pulse energies when the EDF2 with lengths of 7 cm, 10 cm and 20 cm was spliced in the cavity one by one, as shown in Fig. \ref{fig3}, the laser with the three different lengths of EDFs operates in Q-switching regime when the pump power is increased from the lasing thresholds (70 mW, 100 mW and 250 mW for the lengths of EDF2 of 7 cm, 10 cm and 20 cm, respectively) to the maximum 650 mW. The self-started characteristic manifests the effective and high efficiency of the SDIG for Q-switching. From Fig. \ref{fig3} (a), the average powers increase linearly from 0.56 mW, 0.62 mW, 5.83 mW to 27.13 mW, 25.74 mW and 21.38 mW with the gradually raised pump power, and the slope efficiencies are 4.72$\%$, 4.44$\%$ and 3.83$\%$ corresponding to the lengths of EDF2 of 7 cm, 10 cm and 20 cm, respectively. The low slope efficiencies mainly originate from the high loss induced by the narrow reflect bandwidth of the SDIG. Furthermore, during the reduction process of the pump power, the bistable state appears and the obtained minimum emission powers are 0.46 mW, 0.62 mW and 1.05 mW at the pump powers of 63 mW, 80 mW and 120 mW, respectively. During the same modulation process of the pump power, the repeat frequencies and single pulse energies promote while the growth rates reduce [Figs. \ref{fig3}(b) and (d)]. The pulse durations showed in Fig. \ref{fig3}(c) become narrow at first and reach their steady values gradually with the increased pump power. Comparing the results in Fig. \ref{fig3}, we contribute that a larger cavity loss is induced by a longer SDIG due to the large lasing absorption length and narrower bandwidth of the SDIG, leading to the lower average powers, less slope efficiencies and repeat frequencies, larger single pulse energies and broader pulse durations. The deduction is identical with the prediction of the Eq. 3 and Fig. \ref{fig2}(d).

\begin{figure}
	\centering\includegraphics[width=8.5cm]{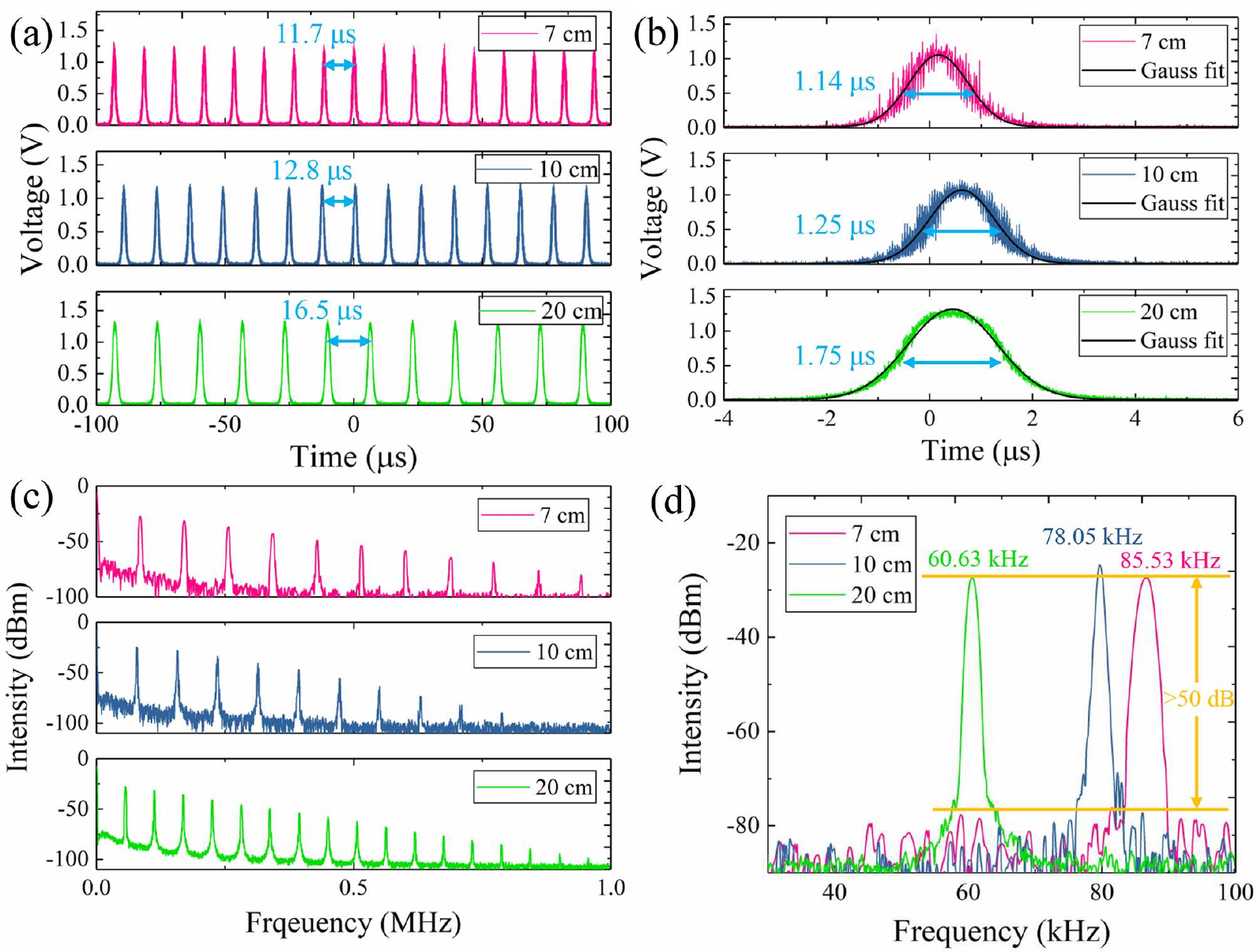}
	\caption{Typical output pulse performance. (a) pulse trains, (b) single pulse waveforms, (c) fundamental frequencies and (d) RF spectrums at the pump power of 450 mW.}
	\label{fig4}
\end{figure}
\begin{figure}
	\centering\includegraphics[width=8.5cm]{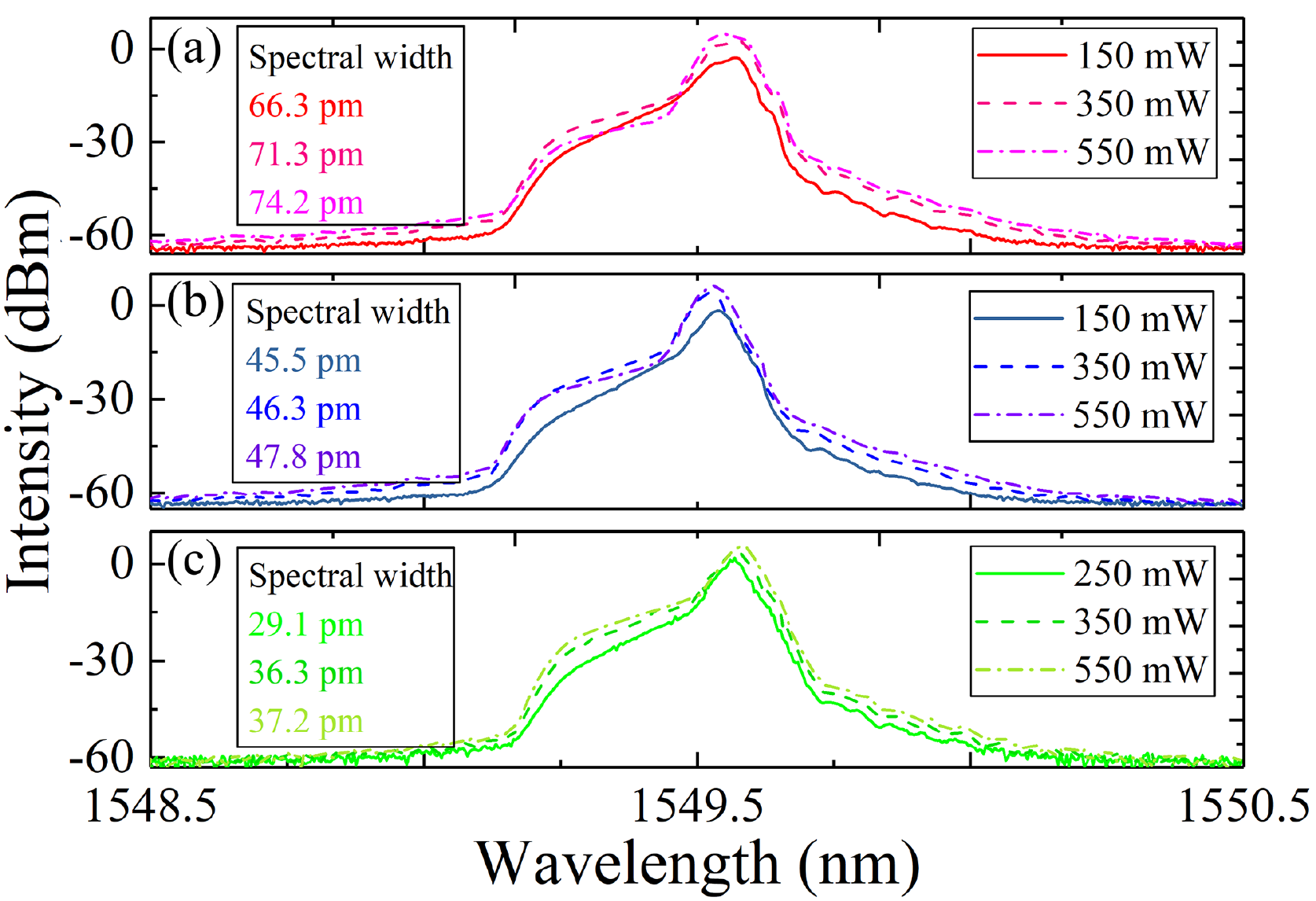}
	\caption{ Optical spectrums of the Q-switched laser with the lengths of the SDIGs of (a) 7 cm, (b) 10 cm and (c) 20 cm, respectively.}
	\label{fig5}
\end{figure}
When the pump power is fixed at 450 mW, the experimental results including the pulse intensities and radio frequency characteristics with the three EDF2 are detected, which are shown in Fig. \ref{fig4}. From Fig. \ref{fig4}(a), the pulse trains in the three lengths of EDF2 are all stable. The pulse intervals of 11.7 $\mu$s, 12.8 $\mu$s and 16.5 $\mu$s correspond to the repeat frequencies of 85.53 kHz, 78.05 kHz and 60.63 kHz in Fig. \ref{fig4}(d), respectively. Figure \ref{fig4}(b) shows the single pulse performances in the expanded time domain. Pulse durations of 1.14 $\mu$s, 1.25 $\mu$s and 1.75 $\mu$s are obtained through Gauss fitting of the pulse data in the three cases. With a shorter EDF2, the noise signal on the Q-switched pulse envelop becomes more obvious, thus the laser tends to unstable. Conversely, acquiring a purer Q-switched pulse needs a longer EDF2. The output performances in the frequency domain are depicted in Figs. \ref{fig4}(c) and (d), the fundamental frequencies (@BW: 2 kHz) manifest the signal-to-noise ratio (SNR) of the three Q-switched pulses exceed 50 dB. Besides, high order harmonic signals exist in the frequency range from 0 to 1 MHz. Obviously, with a fixed pump power, a longer SDIG decreases the repeat frequency and broadens the pulse duration, which is consistent with the results in Fig. \ref{fig3}.
\begin{table*}
	\centering
	\caption{Comparison diagram of the Er-doped fiber lasers based on FSAs ($\sim$ represents the estimated values according to the figures in these literatures).} 
		\begin{tabular}{ccccccc} 
		\hline 
		SAs & Average power (mW) & Repetition rate (kHz) & Central wavelength (nm) & Pulse duration ($\mu$s) & Spectral width (pm) & Refs \\
		\hline 
		\hline
		Tm & - & 0.1-6 & 1570 & 0.42 & - & \cite{oetsai}\\  
		Tm & 100-720 & 0.3-2 & 1580 & 0.1 & $\sim$1100 & \cite{lplkurkov}\\  
		Tm & 0.18-0.24 & 3.9-12.7 & 1557.6 & 7.4-20.6 & - & \cite{cpltiu}\\
		Tm & 136 (Max) & 54.1-106.7 & 1560 & 3.28 & $\sim$200 & \cite{joprahman}\\
		Tm & 1.57 & 14.45-78.49 & 1555.14, 1557.64 & 6.94-35.84 & - & \cite{oclatiff}\\
		Tm-Ho & 0.6-1.1 & 1-15 & 1535-1573 & 8.2-10 & - & \cite{lptao}\\
		Tm-Ho & $\sim$1-12.5 & 5.5-42 & 1557.5 & 7.8 (Min) & 63 & \cite{lpltao}\\
		Tm-Ho & 27.61 (Max) & 11.6-57.14 & 1529.69, 1531.74, 1533.48 & 10.46-61.8 &	- &	\cite{ieeeanzueto}\\
		Sm & 80 (Max) & $\sim$70.2 & 1550.0 & 0.45 (Min) & $\sim$50 & \cite{predaol}\\
		Cr & 10.68 (Max) & 68.12-115.9 & 1558.5 & 3.85 (Min) & $\sim$200 & \cite{oftdutta}\\
		Er & $\sim$2.07 & 0.5-1 & 1530 & 0.08-0.32 & - & \cite{oltsai}\\
		Er & 0.56-27.13 & 17.94-118.79 & 1549.56 & 5.32-1.20 & 29.1 (Min) & This work\\
		\hline
	\end{tabular}
	\label{table1}
\end{table*} 
As to the optical spectra, the variations of shapes and bandwidths concerning the pump power are measured and shown in Fig. \ref{fig5}. The central wavelengths of the optical spectra are around 1549.6 nm, and the shapes remain almost unchanged when the pump power is altered. Due to the bandwidth limitation provided by the SDIGs in EDF2, the full-width-half maximum (FWHM) bandwidth and spectrum structures for each length of the EDF2 broadens with the raising pump power. Besides, when the EDF2 becomes longer, the spectral width obtained from the $\Delta\lambda$ value of the OSA is narrowed significantly. The largest spectral widths are 74.2 pm, 47.8 pm and 37.2 pm with the lengths of the SDIG of 7 cm, 10 cm and 20 cm, respectively. The minimum spectral width of 29.1 pm is obtained when the length of the SDIG and pump power are 20 cm and 250 mW, respectively. The results show that the bandwidth of the SDIG is narrower with a longer EDF2 and lower pump power, which coincide with the theoretical analysis in the section above. Therefore, one can realize a narrower bandwidth and even a single-longitudinal-mode Q-switched pulses with a high power pump source and a piece of longer EDF2.

The pulse characteristics comprising average power, repletion rate, central wavelength, pulse duration and spectral width in several typical published articles on Er-doped fiber lasers Q-switched by different FSAs (including Tm-doped, Tm/Ho-doped, Sm-doped, Cr-doped and Er-doped fibers) are displayed in Table. \ref{table1}. Clearly, the spectral width of this work is the narrowest one in these lasers, indicating the effective filtering of the SDIG. Besides, the tunable ranges of repetition rates and pulse durations of our laser also perform well which are benefit from the compact configuration. 

\section{Conclusion}

We have achieved an Er-doped Q-switched fiber laser with narrow bandwidth pulse emissions based on a self-designed SDIG. Such an FSA based SDIG can provide saturable absorption and spectral filtering simultaneously, which is efficient for realizing Q-switching operation in fiber lasers. Further results manifest that the spectral width of the output Q-switched pulse can be narrowed by increasing the length of SDIG and reducing the pump power. The narrowest spectral width of 29.1 pm is achieved when the SDIG length and pump power are 20 cm and 250 mW, respectively. The theoretical and experimental results are in good agreement. Our method provides a promising way to obtain narrow bandwidth Q-switched fiber lasers with low cost and compact size, which may exhibit significant potentials in nonlinear frequency conversion, Doppler LIDAR and coherent beam combinations.

\begin{acknowledgments}
This work is supported by National Nature Science Foundation of China (61905193); National Key R\&D Program of China (2017YFB0405102); Key Laboratory of Photoelectron of Education Committee Shaanxi Province of China (18JS113); Open Research Fund of State Key Laboratory of Transient Optics and Photonics (SKLST201805); Northwest University Innovation Fund for Postgraduate Students (YZZ17099).
\end{acknowledgments}

\nocite{*}

\bibliography{references}

\end{document}